**Pryde et al. Reply:** In our Letter [1] we propose a scheme for nondeterministic quantum nondemolition (QND) measurement of the polarization of a single photon—a photonic qubit—using linear optics and photodetection. The scheme works with nonunit probability, but success is heralded by the detection of a single photon in the meter output. We provide an experimental demonstration of this scheme and introduce three universally applicable fidelity measures—the measurement fidelity $F_M$, the quantum state preparation fidelity $F_{QSP}$, and the QND fidelity $F_{QND}$—to quantify its performance. The claim of Kok and Munro in their Comment [2] is that one of our fidelity measures $F_M$ is not appropriate because it relies on coincidence measurements. We show why this claim is wrong from both a fundamental and an operational perspective.

The key mistake made by Kok and Munro is to assume that a QND measurement of photon polarization must also be a QND measurement of photon number. This is wrong. A QND measurement of a photonic qubit encoded in polarization need not say anything about the photon number in the signal mode being measured. The Comment states "Coincidence rates imply that we know that two photons were present." A QND measurement of photon polarization does not, and is not required to, measure whether there are two photons present—it measures the polarization of the signal photon, *if* it is present.

The Comment claims that the nonunit efficiency of the single photon detectors used in our experiment requires that our fidelity measure $F_M$ include terms where there is either no photon detected in the meter or no photon in the signal. These outcomes arise due to one of the following: (1) no photon being input into the signal; (2) no photon being input into the meter; (3) a photon being lost in the circuit; (4) an imperfect meter detector registers two photons as one.

It is clear from these possibilities that even if unit efficiency number resolving detectors were used, the measurement would still not tell you whether there was a single photon in the signal output or not—you would have to know there was one there by some other means (i.e., by measuring it). When we detect a single horizontally polarized photon in the meter output we do not "believe [the signal] has a horizontally polarized photon"—we know that if there is a single photon in the signal output it is horizontally polarized with the average probability $F_M$. Practically, (4) is indistinguishable from (1)–(3) without a perfect QND measurement of the photon number at the signal input, or unit efficiency single photon sources, combined with lossless optics.

Note that in our experiment the signal is freely propagating in the sense that we can perform any desired operation on the signal subsequent to the QND measurement (including any number of further photon polarization QND measurements). Quantum information protocols that employ QND measurement typically measure the presence of a photon in the signal mode at the conclusion of the protocol; our measurement fidelity $F_M$ is the appropriate measure for the use of QND measurements in such protocols and that of Kok and Munro is not. For example, consider the quantum key distribution scenario referred to in our Letter in which Eve uses our device to make a QND attack on the line. Eve is only ever interested in her meter results in those situations in which (i) she detects a photon and (ii) Bob detects a photon. She knows which events these are from her own records and from listening in on the public discussion of Alice and Bob after key exchange. These are precisely the events we use to calculate our fidelities, and so our fidelities correctly characterize the effectiveness of her attack. Our measurement fidelity $F_M$ is also the relevant quantity that determines the performance of our QND device in its application to the demonstration of quantum complementarity in our original Letter [1], as well as its application to weak values [3]. Finally, in linear optics quantum information schemes [4], where the photon number is conserved, all protocols ultimately terminate in photon number measurements of all modes; our fidelity $F_M$ applies here as well.

In conclusion, Kok and Munro have wrongly assumed that a QND measurement of the *state* of a qubit must also be a QND measurement of the *presence* of that qubit. Their proposed fidelity measure is inappropriate if one wants to quantify the correlation between the polarization state of a single photon in the signal output and the measurement outcome of the meter; the correct measure is the measurement fidelity $F_M$ proposed in our original Letter. However, if one wanted to quantify the probability of performing a QND measurement of photon polarization *and* a QND measurement of photon number simultaneously, one could use the measure proposed in the Comment, provided unit efficiency single photon sources were available.

G. J. Pryde, J. L. O'Brien, A. G. White, S. D. Bartlett, and T. C. Ralph
  Centre for Quantum Computer Technology
  Department of Physics
  University of Queensland
  Brisbane 4072 Australia

[1] G. J. Pryde *et al.*, Phys. Rev. Lett. **92**, 190402 (2004).
[2] P. Kok and W. J. Munro, preceding Comment, Phys. Rev. Lett. **95**, 048901 (2005).
[3] G. J. Pryde *et al.*, Phys. Rev. Lett. **94**, 220405 (2005).
[4] E. Knill, R. Laflamme, and G. J. Milburn, Nature (London) **409**, 46 (2001).